\documentclass[prb,twocolumn,superscriptaddress,floatfix,showpacs,showkeys,preprintnumbers,amssymb,amsmath]{revtex4}
\usepackage[latin1]{inputenc}
\usepackage{graphicx}
\usepackage{dcolumn}
\usepackage{bm}
\usepackage[mathscr]{eucal}
\usepackage{epsfig}
\usepackage{rotating}

\begin{document}

\newcommand{\beq}{\begin{equation}}
\newcommand{\eeq}{\end{equation}}
\newcommand{\ket}{\rangle}
\newcommand{\bra}{\langle}
\newcommand{\A}{\mathbf{A}}
\preprint{ }
\title{Evidence of Cooper pair pumping with combined flux and voltage control}

\author{Antti O. Niskanen}
\email{antti.niskanen@vtt.fi}
\affiliation {VTT Information Technology, Microsensing, POB 1207, FIN-02044 VTT, Finland}
\affiliation{Low Temperature Laboratory, Helsinki University of Technology, POB 2200, FIN-02015 HUT, Finland}

\author{Jani M. Kivioja}
\affiliation{Low Temperature Laboratory, Helsinki University of Technology, POB 2200, FIN-02015 HUT, Finland}

\author{Heikki Sepp\"{a}}
\affiliation {VTT Information Technology, Microsensing, POB 1207, FIN-02044 VTT, Finland}

\author{Jukka P. Pekola}
\affiliation{Low Temperature Laboratory, Helsinki University of Technology, POB 2200, FIN-02015 HUT, Finland}


\begin{abstract}
We have experimentally demonstrated pumping of Cooper pairs in a single-island mesoscopic structure.
The island was connected to leads through SQUID (Superconducting Quantum Interference Device) loops. 
Synchronized flux and voltage signals were applied whereby the Josephson energies of the 
SQUIDs and the gate charge were tuned adiabatically. 
From the current-voltage characteristics one can see that the pumped current
increases in 1$e$ steps which is due to quasiparticle poisoning on the measurement time scale, but 
we argue that the transport of charge is due to Cooper pairs. 
\end{abstract}


\maketitle

A device that yields a DC current in response to an AC signal at frequency $f$ according to the relation
$I=Qf$ is called a charge pump. In the case of electron pumps $Q=me$ while for Cooper pair pumps $Q=2me$, where 
$m$ is an integer denoting the number of charges being pumped per cycle.
Typically pumping electrons in mesoscopic structures requires an array of at least three
tunnel junctions with voltage gates
coupled to the islands in between the junctions. 
A Cooper pair pump is obtained when the tunnel junctions are replaced by Josephson 
junctions. These devices appear at first sight to be very similar and actually the very same
samples may serve as both Cooper pair and electron pumps depending on whether the device is in the 
superconducting state or not. 
However, major differences exist. 
Besides
the doubled charge in the superconducting state, the nature of the
tunneling processes is very different, too. Electrons can tunnel downhill in energy due 
to the inherent dissipation mechanisms in normal metals with the relevant time scale
given by the $RC$ time constant, where $R$ is the tunnel resistance and $C$ the tunnel capacitance. 
Cooper pairs, on the other hand, try to conserve their energy, and in the absence of an electromagnetic environment,
(i.e. zero impedance) only elastic processes are possible. Their maximum pumping frequency is
proportional to $E_{\rm J}^2/(E_{\rm C}\hbar)$, where $E_{\rm J}$ 
and $E_{\rm C}$ are the Josephson and
charging energies, respectively.
What is more, superconducting circuits may behave coherently in the quantum-mechanical sense.
The first attempt to pump Cooper pairs dates back to over a decade ago \cite{geerligs}. However, 
Cooper pair pumps have not been even nearly as accurate as single-electron pumps. The best example
of the latter ones is the NIST seven-junction pump \cite{7pump}.  The motivation behind pumping Cooper pairs
is two-fold. First of all, Cooper pair pumps are hoped to be able to pump larger currents
than their normal state counterparts while still being accurate. 
This is roughly because increasing $E_{\rm J}^2/(E_{\rm C}\hbar)$ is easier than increasing $1/(RC)$.
Secondly, the operation of Cooper pair pumps
is interesting from the point of view of secondary ``macroscopic'' quantum phenomena and the structures
are quite similar to the superconducting qubits (see, e.g., Refs.~\onlinecite{vion,pashkin}).
Pumping of electrons using surface acoustic waves is another active field of study, see, e.g., Ref.~\onlinecite{saw}.

In this work we report on the experimental demonstration of pumping Cooper pairs
in a structure nicknamed the Cooper pair ``sluice'' introduced and theoretically analyzed 
recently by us, see Ref.~\onlinecite{sluice}.
The device is particularly simple; it has just one superconducting island, like the single Cooper pair transistor,
but the bare Josephson junctions are replaced by SQUID loops. 
The device may be alternatively viewed as a tunable Cooper pair box, a Josephson charge qubit \cite{makhlin}.
Here the control is achieved via adiabatically 
manipulating both the fluxes through the two loops and the gate voltage.
Ideally the SQUIDs act as tunable Josephson junctions whose coupling energy can be varied between 
a value close to zero and the sum of the couplings of the individual junctions.
First we describe the experimental setup and discuss the theoretical idea briefly.
Then we present measured data of the pumping experiment.
We demonstrate that the pumped current obeys nicely the theoretical predictions. 
We also comment on possible ways of improving the results should the device be used in applications
and discuss the significance of the results.

\begin{figure}
\begin{picture}(130,240)
\put(-55,0){\includegraphics[width=0.45\textwidth]{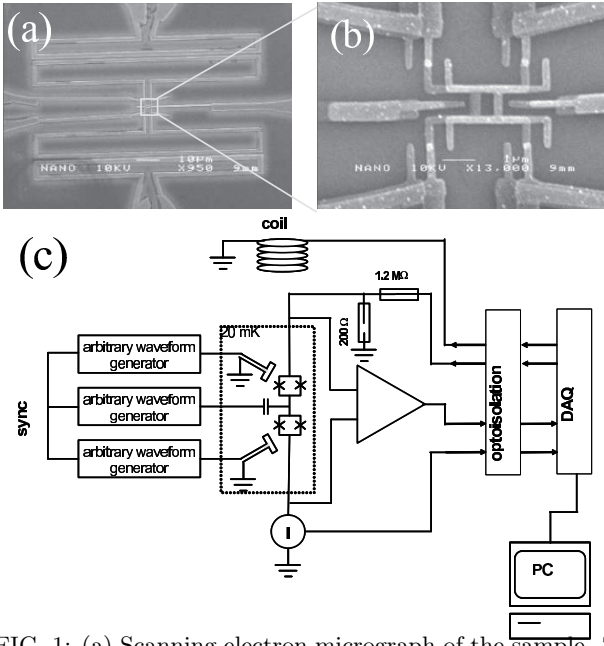}}
\end{picture}
\caption{\label{schematic}(a) Scanning electron micrograph of the sample. 
The two input coils can be seen on top and bottom, respectively.
The gate extends to the far right and the gate capacitance is $C_{\rm g}=0.24$ fF
based on DC measurements.
The current flows between the two leads on the left side. (b) Closeup of the island. The measured total capacitance
of the island is $3.7$ fF which corresponds to a charging energy of about 1 K for Cooper pairs. 
The maximum $E_{\rm J}$ per SQUID is estimated to be 
around 0.5 K based on the normal state resistance. (c) Schematic illustration of the measurement setup. 
We used commercial room temperature
electronics for the current measurement and three synchronized arbitrary waveform generators for the 
control pulse. The external coil for tuning the background of the SQUIDs is at 20 mK. 
The voltage biasing happens via voltage division 
through resistive lines.
A surface mount capacitor of 680 pF and an on-chip capacitor on the order of 10 pF
were also used.}
\end{figure}

Figure~\ref{schematic} shows an SEM image of the sample used in the experiments along with a
schematic of the measurement setup in Fig~\ref{schematic}(c). The device was fabricated out of aluminum 
using standard e-beam lithography and two-angle shadow evaporation. It consists of a
superconducting island that connects to the leads via SQUID loops.
These are relatively large (10 $\mu$m by 100 $\mu$m) in order to have good inductive coupling but 
the island and the junctions are still small such that the charging energy is large enough ($\approx$1 K) to
suppress thermal effects. 
The sample was attached to a dilution cryostat with a base temperature of 20 mK with the RF-lines connected. 

\begin{figure}
\begin{picture}(130,240)
\put(-55,0){\includegraphics[width=0.45\textwidth]{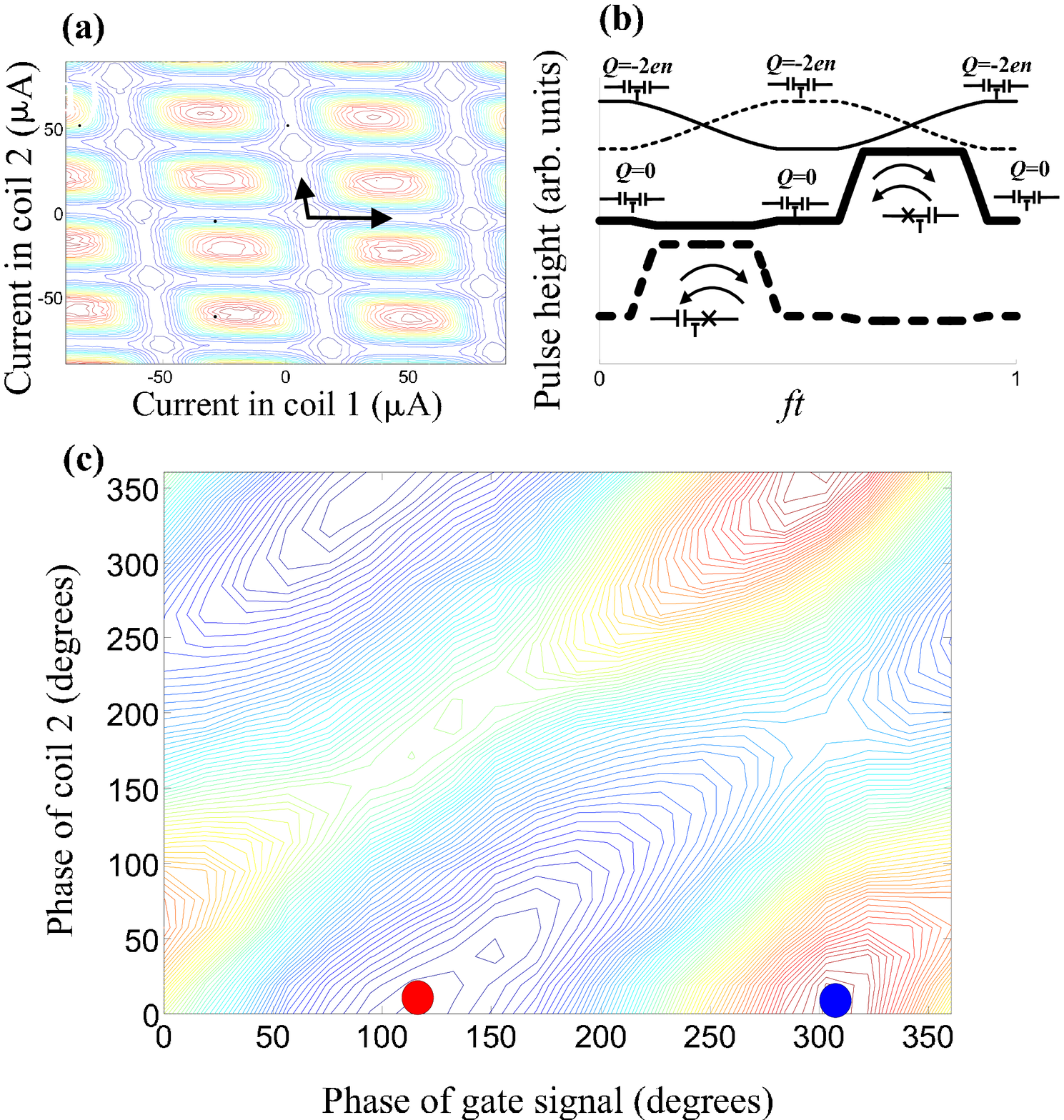}}
\end{picture}
\caption{\label{pulse}(a) Contour plot of the measured DC current at constant voltage against DC currents
in the two input coils. The total variation in the current is around 40 pA at this bias point (150 $\mu$V).
The arrow line indicates the path along which the flux pulsing is performed in the pumping experiment.
The lines of minimum current along which the arrows are aligned are the lines along which
half a flux quantum threads one of the two SQUIDs. The slight tilting of the lines
is a signature of the inductive cross-coupling. Arranging the pulsing as shown compensates for the cross coupling.
(b) Waveforms that were used in the experiment. The thin almost sinusoidal pulse is the gate signal for pumping
in, say, ``forward'' direction, and the dashed $\pi$-shifted signal is for pumping in the ``backward'' direction.
The low level of the gate pulse is zero.
The thick lines are the current signals corresponding to the arrowed path in the previous contour plot.
(c) Contour plot of the 
measured current at a constant voltage of 250 $\mu$V 
against the relative phase differences between the signals
with the pumping signal being applied at 2 MHz. The blue circle is the optimal choice for 
pumping ``forward'' while the red circle is the optimal point for pumping ``backward''. 
The amplitude was set large (over 400$e$) and the variation in current was 150 pA.
This operation point is far from optimal, but we still obtain a clear modulation
for calibration purposes. 
}
\end{figure}

Ideally, the pumping of $m$ Cooper pairs is achieved by applying the three pulses in Fig~\ref{pulse}(b)
through the attenuated RF-lines.
The upmost signal is applied to the gate while the two lower ones represent the currents
flowing in the input coils.  
Two different versions of the gate pulse are shown, one for pumping ``forward'' and one for pumping ``backward''.
To understand how the device works, it is instructive to look at the Hamiltonian of the device,
which reads
\begin{align}\label{ham}
\hat{H}=&E_{\rm C}(\hat{n}-n_{\rm g})^2
-E^{\rm 1}_{\rm J}(\Phi_{\rm 1})\cos(\phi+\varphi/2) \notag \\
&-E^{\rm 2}_{\rm J}(\Phi_{\rm 2})\cos(\varphi/2-\phi).
\end{align}
Here $E_{\rm C}=2e^2/C_\Sigma$ is the charging energy for Cooper pairs where $C_\Sigma$
is the total capacitance seen from the island. Furthermore, $E^j_{\rm J}$ with $j=1,2$ are 
the (signed) Josephson energies of the two SQUIDs which can be tuned with the external fluxes $\Phi_j$. For identical 
junctions $E_{\rm J}^j=E_{\rm J}^{\rm max}\cos(\pi \Phi_j/\Phi_0)$, where $\Phi_0\approx 2\times 10^{-15}$ Wb 
is the flux quantum and $E_{\rm J}^{\rm max}$  
is proportional to the critical current $I_{\rm C}$ of the individual
junctions via $E_{\rm J}^{\rm max}=(\hbar/e) I_{\rm C}$. 
Furthermore, $n_{\rm g}=C_{\rm g} V_{\rm g}/2e$ is the gate charge in $2e$ units, $\hat{n}$ is the number
operator for Cooper pairs, $\phi$ is the phase on the island and their commutator is $[\hat{n},\phi]=i$.
The environment couples to the pump through $\varphi$ which is the phase difference over the pump.
If the SQUIDs were to have perfectly identical junctions as well as vanishing self-inductance  
and if the flux control were perfect then
the effective Josephson couplings could be set to zero. 

Figure~\ref{pulse}(a) shows a contour plot based on the measurement 
of the current through the device at a constant voltage 
against the DC currents in the two input coils. Along the lines of minimum current the flux through
either of the loops is $(k+1/2)\Phi_0$, where $k$ is an integer.
The measurement reveals not only the mutual inductances $M_{ij}$ between coil $i$ and SQUID $j$,
which were $M_{11}=30$~pH, $M_{12}=2$~pH, $M_{21}=3$~pH and $M_{22}=50$~pH,
but also the proper offsets at any given time, 
i.e. the background fluxes threading the loops. This measurement does not fully
demonstrate to which extent it is possible to suppress the Josephson energy. 

In the beginning of an ideal pumping cycle 
the $E_{\rm J}$'s of both loops are set as close to zero as possible 
and the position of the gate determines the ground state. 
We see that initially the ground state of the island is an eigenstate of charge.
We then adiabatically ``open'' one of the SQUIDs, i.e. move to the tip of the, say, horizontal
arrow in Fig~\ref{pulse}(a) which means that the $E_{\rm J}$ of the SQUID 1 is maximized while 
for the other it is still zero. We stay at the tip of the arrow for some time and start to either 
decrease or increase the gate charge $n_{\rm g}$ depending on the direction we have chosen.
When the gate reaches its extremum we ``close'' the SQUID again. Now if everything
has been adiabatic the system is still in its ground state. The charge is again a good quantum number 
at this point but since the position of the gate is different, the number of charges is different too. 
The only possibility 
is that the excess charges have tunneled through the SQUID whose  $E_{\rm J}$ has been non-vanishing during the cycle.
The $E_{\rm J}$ of the second SQUID is then opened and the gate put back to its initial position. Finally the second
SQUID is also closed. The number of Cooper pairs pumped is given by the difference between the integers
closest to the high and low level of the gate charge. Fixing the low level and sweeping the high level
should result in a 2$e$-periodic staircase in the pumped current. 

The phase of the gate determines naturally the direction,
i.e. a 180-degree phase shift reverses the pumped current. 
Fig~\ref{pulse}(c) illustrates the measured behavior of the current when the relative phases between the pulses
are varied. The phase of coil 1 is fixed at 180 degrees and the phases of the other two are swept.  
The two circles shown are the optimal choices for pumping. Note that the extrema of current 
are indeed 180 degrees apart in the gate as expected and the optimal choices are the ones illustrated in Fig.~\ref{pulse}.
For practical reasons we were forced to use frequencies in the MHz
range, but in the present pumping scheme it is possible to increase the 
value of current conveniently by increasing the gate 
amplitude. We tried out different shapes of pulses such as a mere sinusoidal gate signal, but it was found
that it is better to keep the gate constant while the $E_{\rm J}$ is not maximized
which is in accordance with the adiabaticity requirement.
In practice we have arranged for a 15\% dead time between the flux pulses 
although no systematic optimization of the pulses was performed.

\begin{figure}
\begin{picture}(130,100)
\put(-55,0){\includegraphics[width=0.45\textwidth]{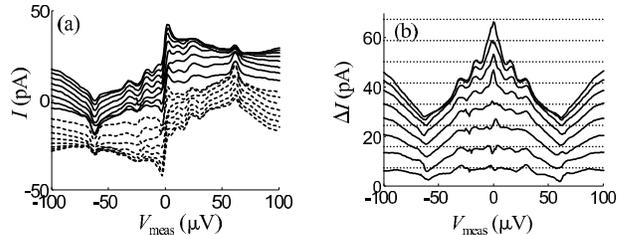}}
\end{picture}
\caption{\label{IV} (a) Examples of measured IV-curves with the pumping signal applied at 3 MHz. 
The gate charge (in $2e$ units) varies
between 4 and 34. The solid curves correspond to pumping forward and the dashed curves correspond to pumping
backward. 
Here $V_{\rm meas}$ is the
measured value of voltage over the pump.
(b) Difference of current, $\Delta I$, in the IV-curves of (a) for pumping in opposite directions. 
The dotted lines indicate the expected values.}
\end{figure}

Figure~\ref{IV}(a) shows an example of characteristic IV-curves (i.e., current-voltage curves) with the pumping signal being applied at 
$f=3$~MHz. The effect of the change of direction is shown. 
The curves correspond to eight different values of gate amplitude.
We see immediately that a leakage current exists on top of the pumped current
that is on the same order or less than the pumped current. The IV-curves, however, clearly shift and the curves 
for pumping in opposite directions are far apart. 
The total current flowing through the device 
is a sum of two contributions,
one being the leakage supercurrent that can be associated with the dynamical phase of the
wave function and the other being the less trivial pumping contribution attributable to
the geometric phase.
If one assumes that the leakage is the same for the pumping
in both directions at a definite voltage bias point, then the difference between the IV-curves should
be twice the magnitude of current pumped in this case.
Fig.~\ref{IV}(b) reveals that at low voltages (tens of $\mu$V) and
at smaller amplitudes this pumping contribution is indeed close to the expected level
shown with dotted lines. 
The leakage current which is due to the nonideal environment and flux control is undesirable from an application point of view, 
but the physical phenomenon is clearly visible. 
The
voltage bias is not sufficiently good to eliminate the leakage, i.e. the P(E)-curve \cite{PE}
for tunneling events is not sufficiently peaked at the origin.

\begin{figure}
\begin{picture}(130,175)
\put(-55,0){\includegraphics[width=0.45\textwidth]{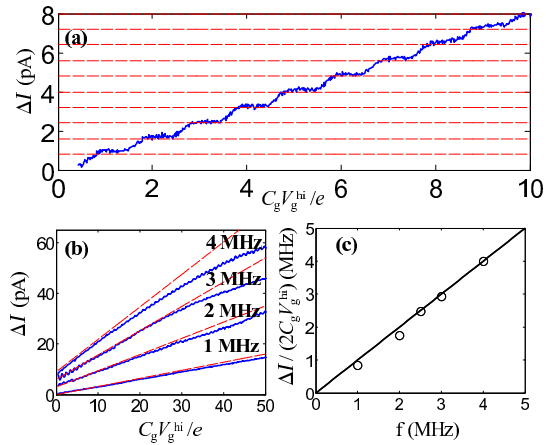}}
\end{picture}
\caption{\label{steps} (a) Difference $\Delta I$ in current of forward and backward pumping at 2.5 MHz against
the high level of the gate signal $V_{\rm g}^{\rm hi}$ with the low level at zero. The dashed lines
are drawn at $2ef$ intervals. 
(b) Large gate amplitude behavior of $\Delta I$ at a few frequencies. 
The dashed lines show the expected gate dependece, i.e. 
their slope is $2ef$. The curves are offset for clarity.
(c) Fitted slopes to the data of the previous plots up to $V_{\rm g} C_{\rm g}/e=10$ are shown by circles. The solid
line indicates the expected behavior. The voltage bias point was around 10 $\mu V$ in all the above plots.
}
\end{figure}

These considerations suggest that it is interesting to study the difference in the currents $\Delta I$
with the gate shifted by 180 degrees. 
Figure~\ref{steps}(a) shows the measured behavior of $\Delta I$
at 2.5~MHz versus the high level of gate voltage with the low level set to zero. 
The current may be seen to increase in clear
steps. The expected height of a step is twice the pumped current, i.e. $4ef$ which in this case is some 1.6 pA. 
Since we sweep the high level of the gate signal and not just the amplitude with constant offset, the 
steps should occur at $2e$ intervals in the gate charge. However, due to random parity changes 
(quasiparticle ``poisoning'') at time scales that are much shorter than our measurement time scale (0.1 s)
but longer than the pumping cycle ($10^{-6}$ s) we observe the time average of two $2e$-periodic
staircases that are shifted by $e$ in the gate charge. For instance in Ref.~\onlinecite{noneq} the
tunneling time for quasiparticles was estimated to be 10 $\mu$s in a similar structure while in
Ref.~\onlinecite{mannik} it was some $10^{-2}$ s for a coupled system of two superconducting transistors with one 
grounded. We were unable to measure the corresponding time in our setup, but
based on this supporting evidence we argue that the 
transport of current is due to Cooper pairs since the order in which the $E_{\rm J}$'s are
manipulated changes the direction of current. The quasiparticles effectively shift 
the gate charge by $e$ but rarely enough such that the pumping is 
undisturbed on the level of precision of the present measurement.
If this interpretation is made then one sees that the obtained results are in 
very good agreement with theory.
Figure~\ref{steps}(b) illustrates the measured large amplitude behavior of the pumped current at frequencies between
1 MHz and 4 MHz. We see that the current lacks behind the prediction with increasing frequency and amplitude.
At 1 MHz no clear bending of the curve is seen up to gate amplitude of 40$e$, while at
4 MHz the performance starts to degrade after $10e$.
One can observe by looking at Fig.~\ref{IV}(b) that the "bending" is more pronounced
at larger bias voltage values (voltage is on the order of 10 $\mu$V in Fig.~\ref{steps}) 
while no visible bending happens up to amplitudes of 68$e$ when $V\approx 0$.
Small amplitude behavior in Fig.~\ref{steps}, however, is linear aside from the steps with a slope of $2ef$. Figure~\ref{steps}(c)
shows the slopes obtained from linear fits to the data of Fig.~\ref{steps}(a) and the ten first
steps of Fig.~\ref{steps}(b). One sees that the agreement is again good.

The above results prove that the flux and voltage driven pumping of Cooper pairs is experimentally 
possible in a single-island device. 
However, in order to serve as a practical device the leakage current 
needs to be taken care of as well as the quasiparticle poisoning.
The quasiparticles may possibly be handled by either quasiparticle ``traps'' or by 
BCS gap profile engineering \cite{noneq}.
As to the reduction of the leakage, several options exist. 
One option is the engineering of the electromagnetic environment such 
that the voltage biasing is good also at frequencies on the order of the charging energy. This would result in 
DC IV-characteristics heavily peaked at zero voltage with negligible leakage current. Another way
to cut down the leakage is to fabricate a longer chain of junctions.
A multiloop SQUID would possibly improve the suppression of $E_{\rm J}$ without increasing the number of controls. 
Improved RF-engineering 
would also be of benefit in arranging the flux pulses. To conclude, the results are encouraging in spite of 
several nonidealities observed and the pumping of Cooper pairs with flux control looks much more attractive than
with a mere multiple gate voltage control.

We thank H. Sipola and S. Franssila for help with the measurement set-up and device fabrication, and
A. Anthore, T. Heikkilä, P. Helistö and M. Paalanen for useful discussions. The
Academy of Finland and EU IST-FET-SQUBIT2 are acknowledged for financial support.

\end{document}